\documentclass[draft]{agujournal2019}
\usepackage{url} 
\usepackage{lineno}
\usepackage{changes} 
\usepackage{soul}

\usepackage{amsmath}
\usepackage{amssymb}

\usepackage{array}
\newcolumntype{C}[1]{>{\centering\arraybackslash}m{#1}}

\journalname{Geophysical Research Letters}
\begin{document}

\title{Amplified Path Variability of the Gulf Stream Extension under Increased Stratification}

\authors{Lennard Miller\affil{1,2}, Antoine Venaille\affil{1}, Stéphane Popinet\affil{3} and Bruno Deremble\affil{2}}

\affiliation{1}{ENS de Lyon, CNRS, Laboratoire de Physique (UMR CNRS 5672), F-69342 Lyon, France.}
\affiliation{2}{Université Grenoble Alpes, CNRS, INRAE, IRD, Grenoble-INP, Institut des Géosciences de l’Environnement, Grenoble, France.}
\affiliation{3}{Sorbonne Universite, CNRS, Institut Jean Le Rond d’Alembert, F-75005 Paris, France.}

\correspondingauthor{Lennard Miller}{lennard.miller@ens-lyon.fr}

\begin{keypoints}
\item A high-resolution ocean model reveals that the warming of the ocean surface amplifies the path variability of the Gulf Stream Extension.
\item The amplified path variability is attributed to enhanced instability of the Gulf Stream Extension under increased stratification.
\item The stratification-driven amplification of path variability is robust to changes in surface winds and meridional overturning.
\end{keypoints}

\begin{abstract}
Increased upper-ocean stratification is an unavoidable consequence of global warming and will strongly impact the structure of ocean currents. Using a high-resolution ocean model, we show that increased stratification amplifies the path variability of the Gulf Stream Extension. This amplification persists independently of changes in the Atlantic Meridional Overturning Circulation and surface wind forcing. A linear stability analysis suggests that the amplified path variability results from enhanced instability growth rates of the Gulf Stream Extension, which in turn are a consequence of the increased shielding of the Gulf Stream Extension from the stabilizing effects of bottom topography as stratification increases. The amplified path variability of the Gulf Stream Extension is not captured by coarse-resolution models that parameterize eddies, and therefore  highlights the need for improved representations of oceanic turbulence in climate projections.
\end{abstract}

\section*{Plain Language Summary}
Oceanic density stratification increases under climate change as surface waters heat and expand. Here we explore how the stratification increase expected by the end of the 21st century alters the dynamics of the Gulf Stream. As stratification increases, we find that the path of the Gulf Stream is distorted by increasingly strong meanders. We further show that this effect holds regardless of whether the Atlantic overturning circulation weakens or winds strengthen, meaning that warmer, more stratified oceans are likely to produce a more meandering Gulf Stream irrespective of these other changes. 
\clearpage

\section{Introduction}

\begin{figure}[htbp!]
    \centering
    \includegraphics[width = \textwidth]{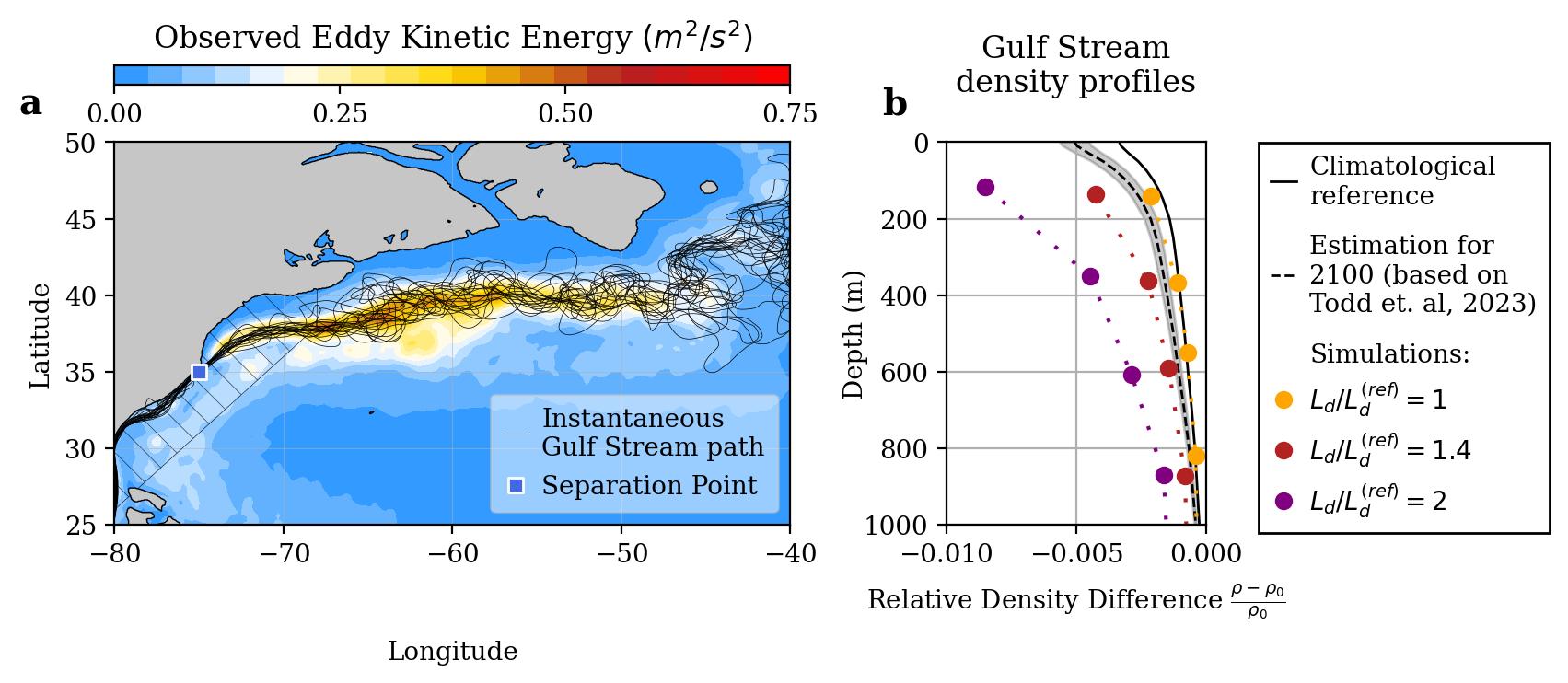}
    \caption{\textbf{Altimetry-derived observations of the GSE (a) and its present, projected and simulated density profiles (b).} The Eddy Kinetic Energy in (a) is calculated from Copernicus Marine Service Altimetry data ~\cite{copernicus-marine-service_2025}. Twenty exemplary instantaneous realizations of the GS path, separated by equal time intervals of ten weeks, are also displayed. In this altimetry data set, the GS path is calculated by tracking the 25cm contour of the sea surface height \cite{andres2016recent, sanchez2024changes}. Downstream of the separation point we refer to the current as the GSE. The GS density profiles shown in (b), defined here as $(\rho-\rho_0)/\rho_0$ with $\rho$ being potential density and $\rho_0$ being a reference density in the abyss, are calculated as spatial averages over the hatched region in (a). The GS region is expected to become more stratified within the next century. Shown here are a climatological reference from~\protect\citeA{levitus1982climatological} and a future estimate, based on data from~\protect\citeA{todd2023warming}, of the density profile along the GS, as well as the profiles in the simulations with climatological AMOC and wind forcing under current stratification ($L_d/L_d^{(\text{ref})} = 1$) and increased stratification ($L_d/L_d^{(\text{ref})} = 1.4$, $L_d/L_d^{(\text{ref})} = 2$). The gray shading shows the uncertainty in the estimated profile. Details on the calculation of the displayed density profiles are found in the Supplementary Text S2.}
    \label{fig: density_prof}
\end{figure}

The Gulf Stream Extension (GSE), the eastward continuation of the Gulf Stream (GS, Fig.~\ref{fig: density_prof}a) beyond Cape Hatteras, plays a central role in North Atlantic climate. Its path anchors the atmospheric jet stream, thereby shaping storm tracks and precipitation patterns across the basin~\cite{kwon2010role, deremble.lapeyre.ea_2012, small.tomas.ea_2014, parfitt2016atmospheric, omrani2019key, larson2024signature}. Additionally, its meanders produce rings that act as hotspots for ocean–atmosphere exchanges~\cite{shaman2010air, gnanadesikan2015isopycnal} with strong ecosystem impacts~\cite{sanchez2015impact, pershing2015slow, haeck2023satellite}. An important aim of oceanic climate science is therefore to anticipate future changes in the variability of the GSE~\cite{fox-kemper.hewitt.ea_2023}.

However, current eddy-resolving climate simulations and observational records are too short to clearly distinguish trends from oscillations in the GSE variability. This has led to an ongoing debate on how to interpret changes in surface current strength~\cite{yang2016intensification, sen2021future, beech2022long, yun2024impact}, mesoscale eddy kinetic energy (EKE)~\cite{martinez2021global, chi2021has, perez2024regime} and the longitude where the GSE destabilizes into meanders~\cite{andres2016recent, sanchez2024changes}.   While oscillations in GSE variability are suggested to be driven by internal ocean dynamics or oscillations in atmospheric forcing, trends have been attributed to several long-term climatic trends across the North Atlantic basin. Surface winds are accelerating~\cite{yang2016intensification, sen2021future}, the Atlantic meridional overturning circulation (AMOC) may be slowing down~\cite{caesar2018observed} and stratification is increasing~\protect\cite{li2020increasing, todd2023warming} as a consequence of the warming of the ocean surface~\protect\cite{wu2012enhanced, pershing2015slow, forsyth2015recent, todd2023warming}. 

The uncertainty and confidence levels associated to each of these climatic trends differ~\protect\cite{fox-kemper.hewitt.ea_2023}. It is thus important to carry out sensitivity studies over a range of possible future scenarios in order to distill the dominant mechanisms that will govern the evolution of the GSE. Idealized model studies have recently shown that changes in stratification can alter the eddies of western boundary current extensions~\protect\cite{miller.deremble.ea_2025}, but the relevance of these findings to the GSE remain unclear as most climate models that run sufficiently long to capture changes in ocean stratification lack the required numerical resolutions to correctly represent the GSE and its eddies (Supplementary Fig.~S3). Therefore, we here focus on the response of the GSE to increased stratification and assess its robustness with respect to changes in AMOC strength and surface wind forcing.

In this article, we present high-resolution ocean simulations which demonstrate that increased ocean stratification amplifies the path variability of the GSE. We carry out a linear stability analysis of the GSE which suggests that this amplification of path variability is a consequence of an enhanced instability of the GSE under increased stratification. We further show that this stratification-driven amplification is robust across different boundary conditions, persisting when the AMOC is suppressed or when wind forcing is increased beyond climatological values.

\section{The Isolated Response of the Gulf Stream Extension to Increased Stratification}

\begin{figure}[htbp!]
    \centering
    \includegraphics[width = 0.5\textwidth]{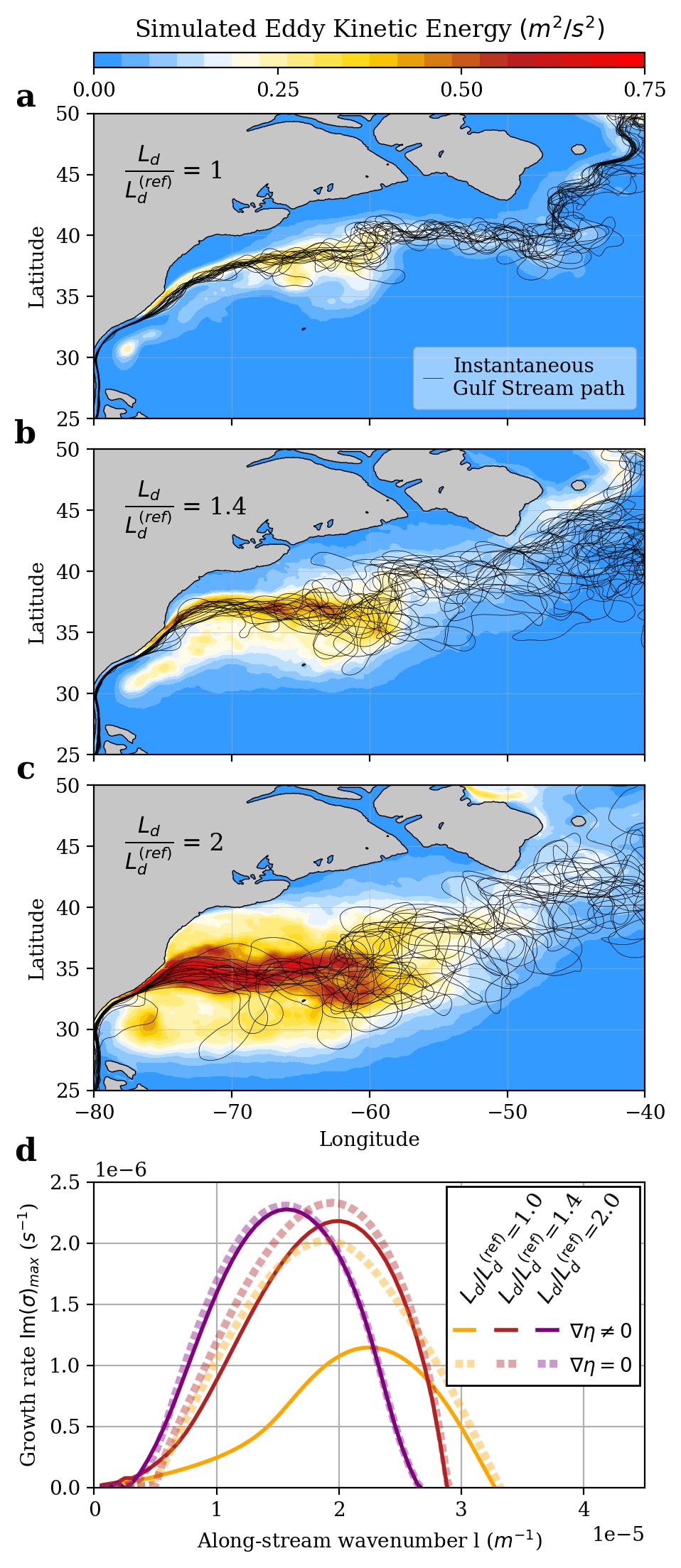}
    \caption{\textbf{EKE (color scale) and path lines of the GSE under climatological AMOC and wind forcing at $\mathbf{L_d/L_d^{(\text{ref})}~=~1}$ (\textbf{a}), $\mathbf{L_d/L_d^{(\text{ref})}~=~1.4}$ (\textbf{b}) and $\mathbf{L_d/L_d^{(\text{ref})}~=~2}$ (\textbf{c}), and associated maximum instability growth rates $\operatorname{\mathbf{Im}}(\boldsymbol{\sigma})\mathbf{_{max}}$ (d).} In our model the GS path is tracked by the 0m contour of the sea surface height. In each panel, 20 instantaneous path realizations spaced by equal time intervals of 10 weeks are shown. Their spread grows larger under increased stratification, illustrating the amplified path variability of the GSE. EKE also increases under stronger stratification. The growth rates displayed in (d) compare the results of a stability analysis of an idealized model of the GSE (Supplementary Text S4) where topographic variations are either taken into account ($\nabla\eta \neq 0$) or not ($\nabla\eta = 0$). The maximum growth rate of the full stability analysis ($\nabla\eta\neq 0$) increases under increased stratification because the stabilizing effect of topography on the GSE under reference stratification ($L_d/L_d^{(\text{ref})}=1$) vanishes in the increased stratification scenarios ($L_d/L_d^{(\text{ref})}=1.4,2$).}
    \label{fig:fluctuations}
\end{figure}

We investigate the impact of increased stratification on the GSE using a high-resolution isopycnal ocean model of the North Atlantic. This model requires the user to specify the basin-averaged depth and the density of the isopycnal layers, which is equivalent to prescribing the volume and density of water masses that are present in the North Atlantic basin. The climatological reference values for the average layer depths and layer densities were chosen following~\citeA{hurlburt2000impact}, and agree well with spatial averages of the Levitus climatology over the GS area (Fig.~\ref{fig: density_prof}b). Forced by climatological wind stress~\cite{hellerman1983normal} and AMOC mass fluxes at the zonal domain boundaries, this reference configuration of the model correctly reproduces spatial variations in stratification across the basin (Supplementary Fig.~S8), the observed present-day surface circulation and AMOC structure (Supplementary Fig.~S1, S2) and a GSE flowing from Cape Hatteras to the Great Banks as a meandering, unstable jet (Fig.~\ref{fig:fluctuations}a).

Motivated by projections~\protect\cite{wu2012enhanced} and observations~\protect\cite{li2020increasing, todd2023warming} of decreases in surface density under global warming, we study the isolated response of the GSE to increased stratification by decreasing the density of the surface isopycnal layers in our model while applying the same climatological AMOC and wind forcing (Supplementary Text~S2). Two simulations were carried out where the layer density deficits $\rho_i - \rho_0$ ($i$ being the layer index with $i=0$ denoting the abyssal layer) were multiplied by $2$ and $4$. The multiplication by a constant factor mimics that ocean stratification profiles as well as the observed changes thereof are surface-intensified~\cite{todd2023warming}. We quantify the model stratification by calculating a basin-averaged value of the first baroclinic Rossby radius of deformation $L_d$. The deformation radii in the investigated scenarios range from a climatological reference value of $L_d^{(\text{ref})} = 32$ km to doubled values, $L_d/L_d^{(\text{ref})}=2$. Fig.~\ref{fig: density_prof}b shows that the intermediate case ($L_d/L_d^{(\text{ref})}=1.4$) describes a scenario in which model stratification approximates an extrapolation of the currently observed GS stratification trends~\cite{todd2023warming} to the end of the 21$^{\text{st}}$ century. The extreme case $L_d/L_d^{(\text{ref})} = 2$ serves to further demonstrate the increase in GSE path variability by reinforcing the system response to stronger stratification, analogous to extreme increases in atmospheric $CO_2$ content commonly applied in earth system models to reveal climatic trends~\cite{andrews.gregory.ea_2012}. In all our model configurations, we fully resolve mesoscale eddies as the grid spacing of our model (5~km, 0.04°) is always smaller than $L_d$.

\subsection{Amplified Path Variability of the Gulf Stream Extension}

The GSE path, with its latitude and longitude denoted as $\theta_\text{GSE}$ and $\phi_\text{GSE}$, is tracked closely by the 0m sea surface height (SSH) contour in our model ($\text{SSH}(\theta_\text{GSE}, \phi_\text{GSE}) = 0$m, Fig.~\ref{fig:fluctuations}a). Fig.~\ref{fig:fluctuations}b and c show qualitatively that the variability of the GSE path increases under increased stratification scenarios (see also Supplementary Movie S1). At $L_d / L_d^{(\text{ref})} = 1.4$, larger meanders induce a significant spread downstream of the detachment point, resulting in a notable increase of GSE path variability east of 60° W. At $L_d / L_d^{(\text{ref})} = 2$ detachment latitude varies and frequent retroflections appear (Fig.~\ref{fig:fluctuations}c), suggesting that increased stratification can lead to an upstream displacement of the point where the GSE destabilizes into meanders \cite{andres2016recent, sanchez2024changes}. 

We quantify the path variability by calculating the standard deviation $s_\text{GSE}$ of the latitude $\theta_\text{GSE}$ as
\begin{equation}
    s_\text{GSE}(\phi) = \sqrt{ \ \langle\left(\theta_\text{GSE} - \langle\theta_\text{GSE}\rangle^\phi\right)^2\rangle^\phi \ }.\label{eq: GSE_path_variability}
\end{equation}
Here $\langle\cdot\rangle$ denotes a GSE path average, calculated as an average along the GSE path and in time (Supplementary Text S3). The superscript $\langle\cdot\rangle^\phi$ denotes a path average across all GSE path segments with $\phi<\phi_\text{GSE}<\phi + 1$°, and thus implies that the path variability $s_\text{GSE}$ is calculated as a function of $\phi$. Fig.~\ref{fig:profiles}a quantitatively compares the simulated GSE path variability to observations, calculated by tracking the contour $SSH = 0.25m$~\cite{sanchez2024changes} in Copernicus altimetry data~\cite{copernicus-marine-service_2025} over the period from June 2014 to June 2024. The GSE path variability increases everywhere between 80° W and 50° W as stratification increases, with $s_\text{GSE}$ at $L_d/L_d^\text{ref}=2$ exceeding twice the values of the reference simulation and observations. 

\subsection{Increase of Eddy Kinetic Energy}

In conjunction with the enhanced GSE path variability we also observe an increase in EKE, defined as 
\begin{equation}
    \text{EKE} = \frac{\overline{|\mathbf{u} - \overline{\mathbf{u}}|^2}}{2}
\end{equation}
with $\mathbf{u}$ being the surface velocity, $|\cdot|$ the vectorial magnitude and $\overline{\cdot}$ the time-average. The increase in EKE is visible in the color scale of Fig.~\ref{fig:fluctuations}. A more quantitative comparison is provided in Fig.~\ref{fig:profiles}e, which presents the zonally averaged EKE profiles calculated as
\begin{equation}
    \overline{\text{EKE}}^\phi = \frac{1}{\phi_2 - \phi_1}\int_{\phi_1}^{\phi_2}\ \text{EKE} \ d\phi
\end{equation}
with $\phi_1 = 80$~°W and $\phi_2=40$~°W. With respect to the reference case, the peak values of the zonally averaged EKE profiles grow roughly as the square of $L_d/L_d^{(ref)}$ (Fig.~\ref{fig:profiles}e). Additionally the latitudinal extend of the EKE pattern grows, visible in the wider spread of yellow areas in Fig.~\ref{fig:fluctuations}b and c and the widening of the EKE profiles in Fig.~\ref{fig:profiles}e. 

\subsection{Mechanism: Weakened Coupling between the Gulf Stream Extension and Topography}

Topography has long been suggested to exert a dynamical control on the GSE \protect\cite{thompson1971there,dalibard2026linear}, supported by the celebrated locking of the Gulf Stream path onto bathimetric contours \protect\cite{wang2017decomposition}. Since increased upper-layer stratification implies stronger screening of surface currents from the deep ocean~\protect\cite{wang2024more}, topography should play a lesser role in shaping the GSE when stratification increases. This is confirmed in our numerical experiments by the southward shift of the GSE (Fig.~\ref{fig:fluctuations}) and its EKE signature (Fig.~\ref{fig:profiles}e) as $L_d/L_{d}^{\mathrm{(ref)}}$ increases.

The amplified variability of the GSE path can also be understood as a consequence of the decoupling of the GSE from topography under increased stratification. Topography has been suggested to exert a stabilizing effect on perturbations of the GSE~\cite{lacasce2019jet, flierl2026topographic}, and we demonstrate here that this stabilizing effect vanishes in our simulations under increased stratification. To this end, we carry out a linear stability analysis of the GSE (Supplementary Text S4) under the assumption that the current develops a localized instability after its detachment from the coast~\cite{dimas1995baroclinic}. This allows representing the GSE in the stability analysis as a current that does not vary along its flow direction. We further simplify the analysis by approximating the GSE velocity profile with a Gaussian shape function that is fitted to the path averaged velocity profiles of the GSE in our numerical simulations (Supplementary Text S3, Table S2, Fig.~S4). In order to calculate the evolution of small perturbations around the GSE, we use a quasi-geostrophic model that is tuned based on path averages of planetary vorticity and topography along the GSE (Supplementary Text S3, Table S1). The perturbations are decomposed into Fourier modes along the GSE flow direction and in time, and the derivatives orthogonal to the flow direction are discretized. The resulting eigenvalue problem for the complex temporal frequency $\sigma$ is solved numerically with python scripts that are linked in the open research section of this article.

Following this methodology, we perform a linear stability analysis of the GSE for each of the three stratification scenarios. The solid lines in Fig.~\ref{fig:fluctuations}d show the maximum instability growth rates $\operatorname{Im}(\sigma)_{max}$ as a function of along-stream wavenumber $l$ that are obtained when topographic gradients are taken into account ($\nabla\eta \neq 0$). The overall maximum value of $\operatorname{Im}(\sigma)$ across all wavenumbers more than doubles as stratification increases from $L_d/L_d^\text{ref} = 1$ to $L_d/L_d^\text{ref} = 2$, in accordance with the amplified variability of the GSE path observed in the comprehensive simulations.

We attribute the increase in growth rate to a decoupling of the GSE from topography. To support this claim, we characterize the sensitivity of the GSE stability to topography by recalculating the growth rates when topographic gradients are not taken into account ($\nabla\eta = 0$). The corresponding dashed curves of $\operatorname{Im}(\sigma)_{max}$ in Fig~\ref{fig:fluctuations}d do not change significantly across the analyzed stratification scenarios, suggesting that changes of the GSE in shape and speed (Supplementary Fig.~S4) have an overall negligible effect on its stability. With respect to these test cases, accounting for topographic gradients almost halves the growth rates under reference stratification ($L_d/L_d^\text{ref}=1$) but does not significantly affect the growth rates at increased stratification ($L_d/L_d^\text{ref}=1.4,2$). The stabilizing effect of topography on GSE perturbations is thus shown to vanish as stratification increases.

\section{Robustness of Amplified Path Variability With Respect to AMOC and Wind Changes}

\begin{figure*}[htbp!]
    \centering
    \includegraphics[width = \textwidth]{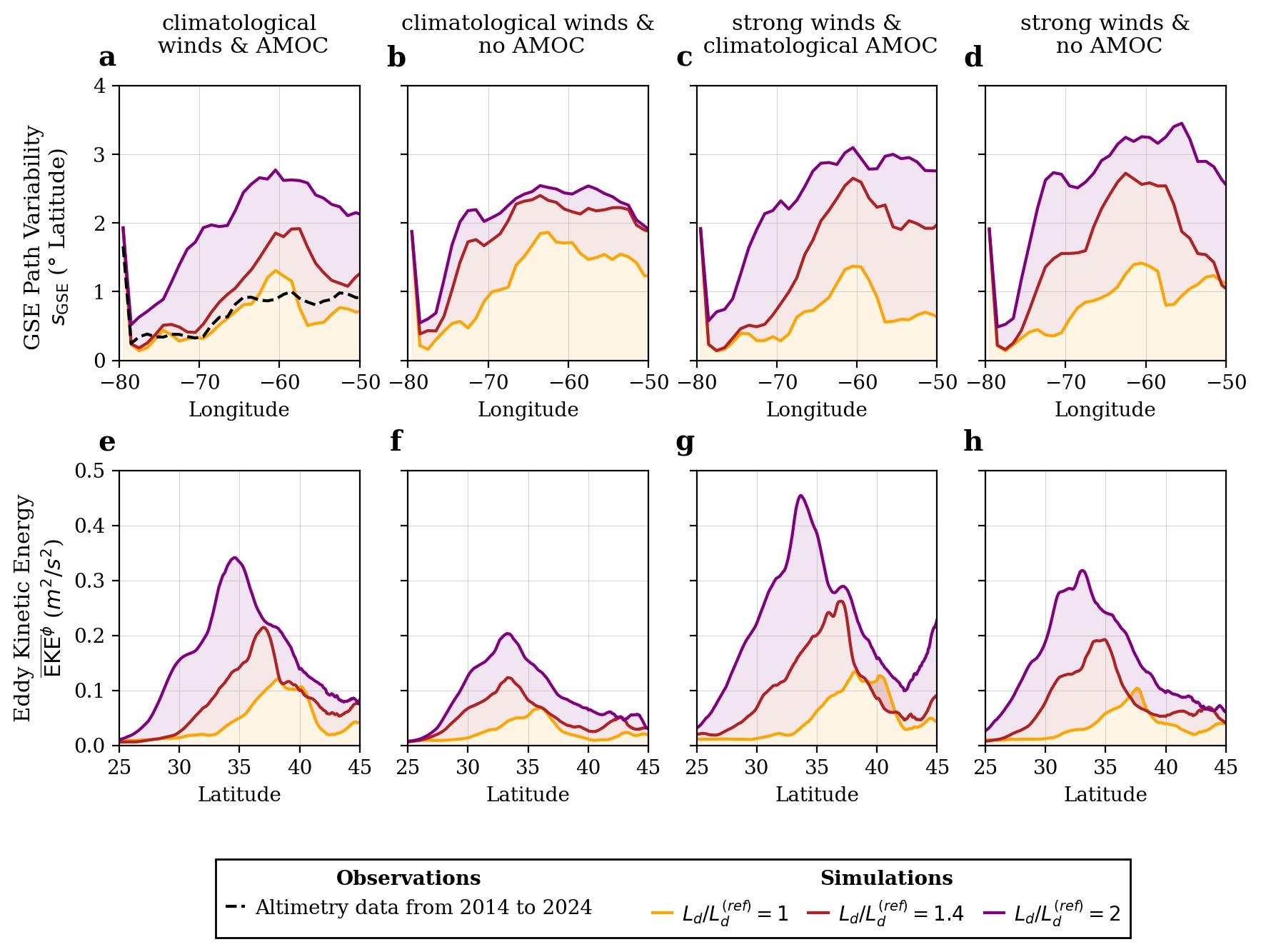}
    \caption{\textbf{Robustness of the stratification-driven increase in GSE path variability and EKE across different forcing scenarios}. Top panels (a, b, c, d) show GSE path variability ($s_\text{GSE}$) as a function of longitude. Bottom panels (e, f, g, h) show zonal averages of eddy kinetic energy ($\overline{\text{EKE}}^\phi$) as a function of latitude. Each column corresponds to a different combination of wind and AMOC forcing: climatological winds and AMOC (a, e), climatological winds with no AMOC (b, f), stronger winds with climatological AMOC (c, g) and stronger winds with no AMOC (d, h). Within each column, results are shown for three stratification levels: reference ($L_d/L_d^{(\text{ref})} = 1$, orange), intermediate ($L_d/L_d^{(\text{ref})} = 1.4$, red) and doubled ($L_d/L_d^{(\text{ref})} = 2$, purple).  In all cases, both path variability and EKE increase monotonically with stratification. Observed path variability, derived from surface altimetry data covering the period from June 2014 to June 2024, is included as a black dashed curve in panel a.}
    \label{fig:profiles}
\end{figure*}

The variability of the GSE can change not only due to increased stratification, but also as a response to the changes in volumetric transport of the GS~\cite{beech2022long} following increases in surface winds~\cite{yang2016intensification, sen2021future} and a slowing of the AMOC ~\cite{caesar2018observed}. In order to investigate the impact of these climatic trends, we perform a sensitivity study by running three additional sets of simulations in which (i) the AMOC is suppressed entirely by removing the mass flux boundary conditions (Supplementary Fig.~S5), (ii) the surface wind stress is increased by a factor of 1.5 beyond the climatological values of~\citeA{hellerman1983normal} (Supplementary Fig.~S6) and (iii) both the AMOC is supressed and wind stress is increased (Supplementary Fig.~S7). Each set is composed of simulations describing the same three increasing stratification scenarios used in the configurations with climatologial AMOC and wind forcing. 

Fig.~\ref{fig:profiles} summarizes GSE path variability and EKE as a function of stratification for all three sets of boundary conditions. In every case, both path variability (top panels) and EKE (bottom panels) increase monotonically with stratification, qualitatively reproducing the behaviour of the simulations with climatological winds and AMOC. The absence of the AMOC (Fig.~\ref{fig:profiles}b, f) decreases the baseline levels of EKE but does not suppress the stratification-driven amplification of GSE path variability. Similarly, stronger winds (Fig.~\ref{fig:profiles}c, g) raise the overall EKE of the system but leave the qualitative ordering with increased stratification unchanged. The enhancement of GSE path variability under increased stratification is thus robust to these other factors that are anticipated to alter GSE dynamics under climate change.

\section{Discussion}

Most climate models lack the horizontal resolution to resolve the GSE~\cite{kirtman.bitz.ea_2012,fox-kemper.hewitt.ea_2023}. Consequently, they cannot capture the stratification-driven amplification of GSE path variability we identify. We also performed low-resolution simulations which illustrate the negligible impact of changing stratification on the structure of the GSE in that case (Supplementary Fig.~S3). This highlights the necessity to either employ eddy-resolving ocean components in climate models or to develop novel parameterizations that capture the behavior of the GSE under varying stratification.

Alongside enhanced GSE path variability we also observe that increased stratification leads to increases in EKE. The increases in EKE peak values (Fig.~\ref{fig:fluctuations}) are likely related to the higher flow speed of the GS under increased stratification (Supplementary Fig. S4, \protect\citeA{peng2022surface}). The latitudinal widening of the EKE profiles in Fig.~\ref{fig:profiles} could be a signature of the enhanced path variability of the GSE but also of larger GS rings that form in our simulations under increased stratification (Supplementary Movie S1). While an increase in ring size is in qualitative agreement with the increase in $L_d$ and the decrease in the most unstable wavenumber of the GSE (Fig~\ref{fig:fluctuations}d), further research is required to quantify and understand the impact of increased stratification on GS eddies and on their non-linear interaction with the GSE.

Recent studies have linked increased ocean stratification under anthropogenic climate change to stronger surface currents~\cite{peng2022surface}, weaker abyssal flows~\cite{wang2024more} and lateral shifts in western boundary currents~\cite{yang2025onshore}. This article further extends the literature on the dynamical impact of increased stratification by revealing that anticipated changes in stratification over the 21st century amplify the path variability of the GSE. An important avenue for future research is to investigate whether the extensions of other western boundary currents such as the Kuroshio or the Agulhas current respond similarly to increased stratification.

While we observe the meridional structure of the AMOC to remain stable in our simulations under increased stratification (Supplementary Fig.~S2b–d), our modeling approach may miss possible feedback of increased eddy activity onto the magnitude of AMOC transport or oceanic stratification~\cite{li2022drivers}. It is also possible that the dynamics of the GSE are sensitive to spatial variations in stratification changes that result from the heterogeneity of the underlying buoyancy fluxes. However, we do not explicitly represent these fluxes in our hydrodynamical model. Our model also lacks the vertical resolution to simulate wind-driven changes in the mixed layer depth~\protect\cite{treguier2025multidecadal}, which could alter submesoscale variability~\cite{callies2016role}. Further sensitivity studies with fully-coupled eddy-resolving climate simulations~\cite{beech2022long, yun2024impact} appear to be a promising tool to better understand both the effect of heterogeneous ocean warming and climatic feedback loops in order to increase fidelity in quantitative predictions for the dynamical changes of the GSE.

\subsection{Conclusion}

We have shown that increased upper-ocean stratification, as projected for the end of the 21st century, will amplify the path variability of the GSE. This amplification occurs because increased stratification gradually decouples the GSE from the stabilizing effects of topography. The amplification of GSE path variability and associated increases in EKE are shown to be robust responses to increased stratification that persist independently of changes in the AMOC and surface wind forcing. 

The amplification of GSE path variability is not captured by coarse-resolution climate models that do not resolve the GSE, underscoring the need for eddy-resolving ocean components or improved mesoscale parameterizations in future climate projections. The associated increase in the variability of air-sea fluxes could potentially increase the latitudinal spread of North Atlantic storm tracks~\cite{deremble.lapeyre.ea_2012, small.tomas.ea_2014, feng.huang.ea_2017}, calling for future work that assesses impacts on weather patterns in the northern hemisphere.

\clearpage

\section*{Open Research Section}

The code for the ocean circulation model employed to produce the raw simulation outputs can be accessed freely online (https://basilisk.fr/src/examples/gulf-stream.c, accessed on the 30th of June 2026). The figures of this article can be reproduced with jupyter notebooks that are stored on Zenodo~\cite{Miller2026_Destabilization_Figures}. The Levitus World Ocean Atlas data \cite{levitus1982climatological} used to calculate the reference Gulf Stream stratification profile of figure 1 and the ETOPO2 dataset~\cite{noaa2022etopo} that contains the bathymetry used in the ocean model can be obtained from the NOAA PSL, Boulder, Colorado, USA, from their website at https://psl.noaa.gov. The Hellermann wind climatology which is used to force the ocean circulation model is distributed online by NCAR~\cite{hellerman1983normal}. The AMOC observation data from the RAPID array used in figure S2 are freely available from https://rapid.ac.uk/. The DUACS altimetry data used in figure S1 is provided on request by the Copernicus program through the Climate Data Store~\cite{copernicus-marine-service_2025}.

\section*{Conflict of Interest declaration}
The authors declare there are no conflicts of interest for this article.

\acknowledgments
This project has received financial support from the CNRS through the 80 Prime program and was also partially funded by the French Agence Nationale de la Recherche (ANR), under the projects ANR-23-CE01-0009 and ANR-23-EXMA-0003. The numerical simulations were performed using HPC resources at PSMN Lyon and GENCI HPC (allocation A0150112020). Data from the RAPID AMOC observing project is funded by the Natural Environment Research Council, U.S. National Science Foundation (NSF) with support from NOAA.

\nocite{popinet2020vertically, moat2024atlantic,deremble_qgutils, lacasce2019jet, levitus1982climatological,vallis2017atmospheric}

\bibliography{biblio}

\newpage




\setcounter{figure}{0}
\renewcommand{\thefigure}{S\arabic{figure}}
\noindent{\textbf{Supplementary Information for "Amplified Path Variability of the Gulf Stream Extension under Increased Stratification"}}

\vspace{20mm}

\noindent\textbf{Contents of this file}

\begin{enumerate}
\item Texts S1 to S4
\item Tables S1 and S2
\item Figures S1 to S8
\end{enumerate}
\noindent\textbf{Additional Supporting Information (Files uploaded separately)}
\begin{enumerate}
\item Captions for Movie S1
\end{enumerate}

\clearpage

\noindent\textbf{Introduction}\\

\begin{figure}[htbp!]
     \centering
     \includegraphics[width = \textwidth]{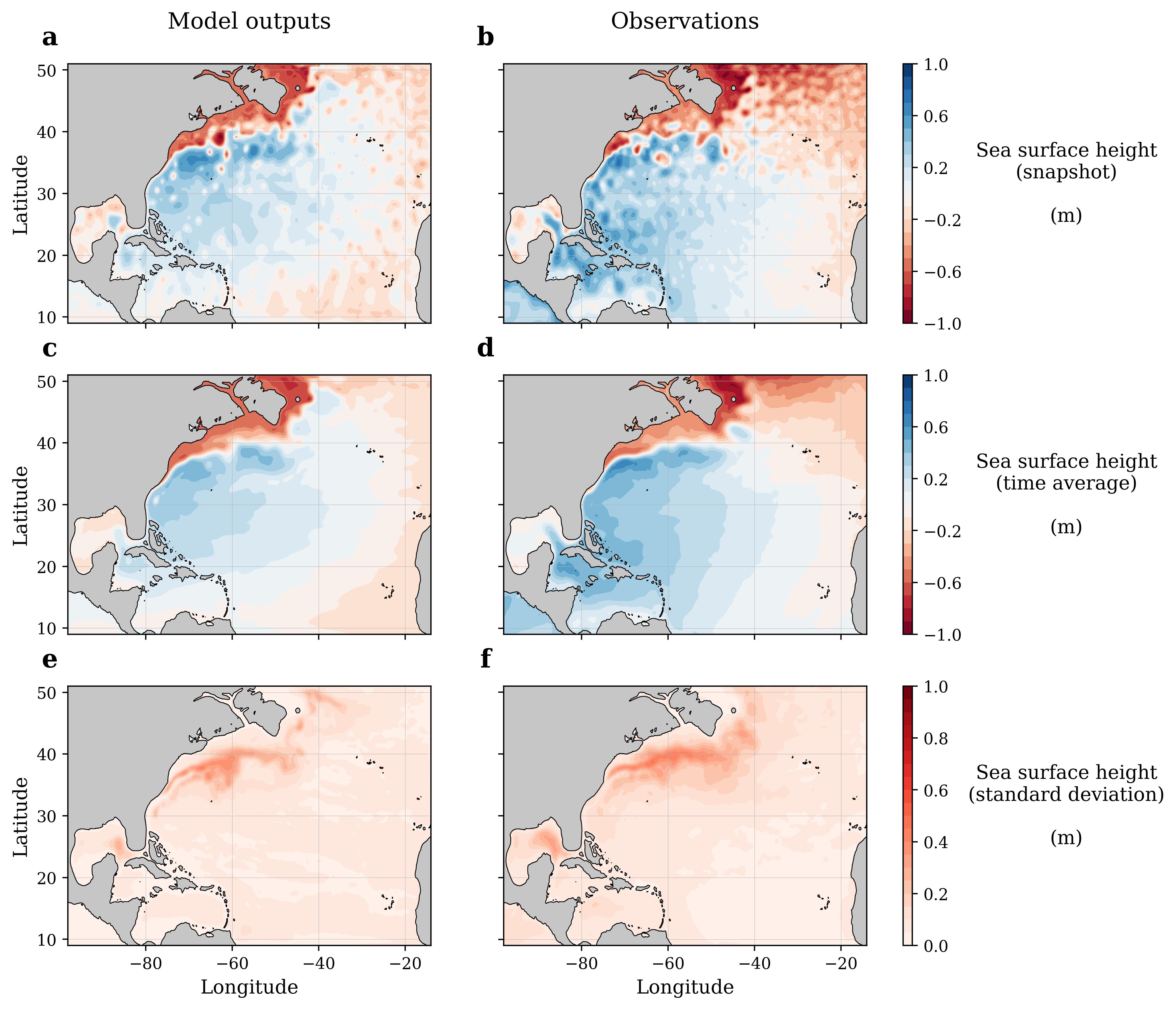}
     \caption{\textbf{Comparison of sea surface height  between model outputs at reference conditions ($\mathbf{L_d/L_d^{(\text{ref})}~=~1}$, left column) and observations from Copernicus Marine Service Altimetry data (right column)~\cite{copernicus-marine-service_2025}.} The top row (\textbf{a}, \textbf{b}) shows instantaneous snapshots, the middle row (\textbf{c}, \textbf{d}) time average fields  and the bottom row (\textbf{e}, \textbf{f}) the standard deviation  of the sea surface height. For increased comparability the snapshots and time-averaged fields have been offset by the domain average of SSH. The model successfully reproduces both the large-scale gyre signature and the variability and structure of the GSE.}
     \label{fig: validation_AVISO+}
 \end{figure}

\begin{figure}[htbp!]
     \centering
     \includegraphics[width = \textwidth]{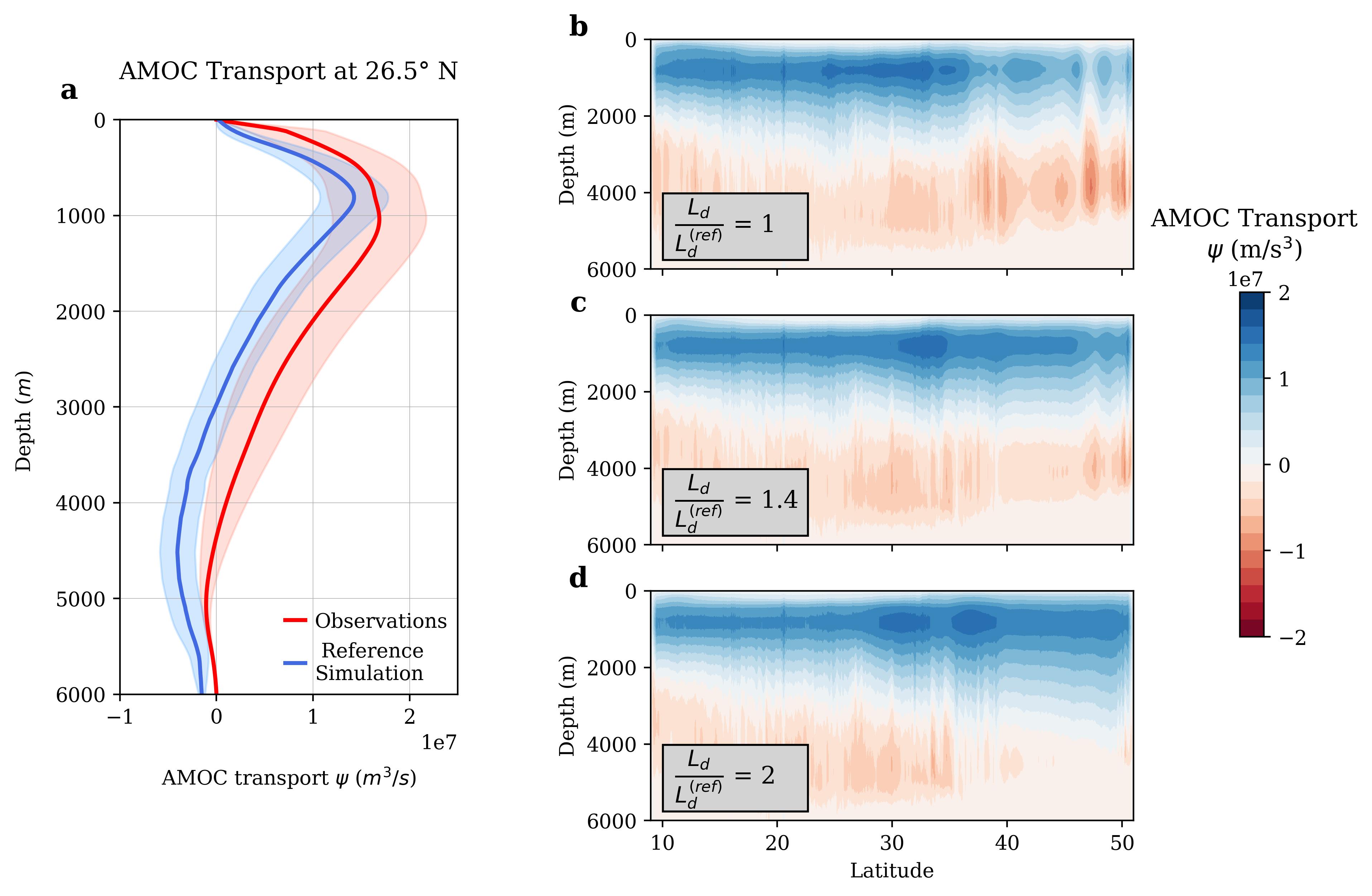}
     \caption{\textbf{Model Representation of the Atlantic Meridional Overturning Circulation (AMOC).} The AMOC transport is defined as  $\psi(y,z) = \iint v(x,y,z') \ dz' \ dx$, where the $x$-integral is carried out across the entire oceanic basin and the $z'$-integral from the ocean surface down to a height $z$. Both the time-averaged transport as well as the range of temporal variability (standard deviation, shaded regions) of the model under reference stratification ($L_d/L_d^{(\text{ref})}$ are compared against observations at 26.5°~N in \textbf{a}. The observational data is derived from a 19-year time-series measured by the RAPID array~\cite{moat2024atlantic}. The meridional structure of the model overturning transport $\psi$ is shown in \textbf{b} for $L_d/L_d^{(\text{ref})} = 1$, in \textbf{c} for $L_d/L_d^{(\text{ref})} = 2$ and in \textbf{d} for $L_d/L_d^{(\text{ref})} = 4$. Water is transported along the contour lines of $\psi$, and the color scale denotes the transported water volume. Despite the amplified path variability of the GSE at increased stratification the AMOC does not change significantly, illustrating that changes in GSE variability do not necessarily provoke a response in the AMOC.}
     \label{fig: AMOC}
 \end{figure}

\begin{figure}
     \centering
     \includegraphics[width = 0.45\textwidth]{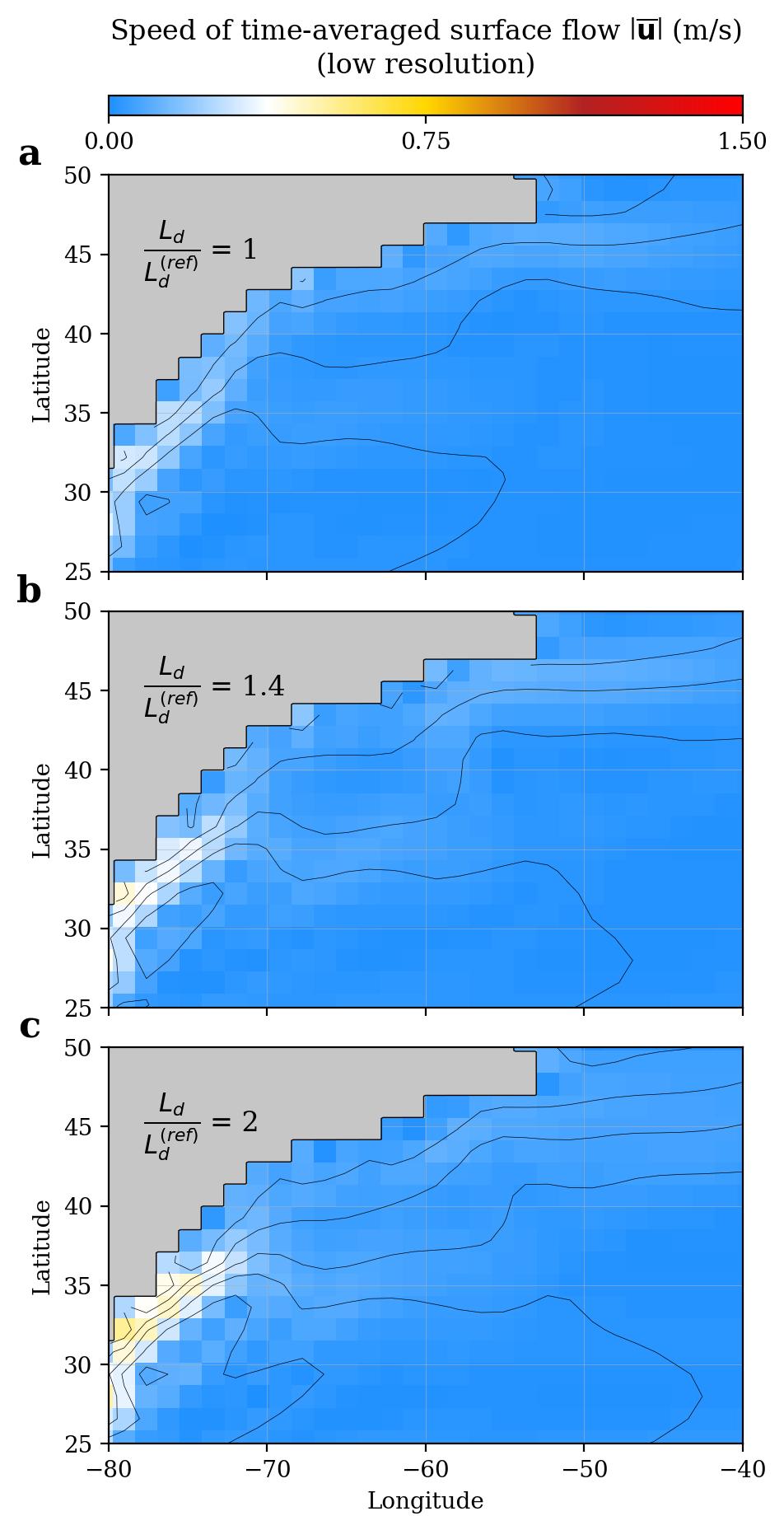}
     \caption{\textbf{Time-averaged GS under climatological winds and AMOC forcing at low numerical resolution for increasing stratification.} The speed of the time-averaged surface flow is shown by the color scale, and the black contours show lines of constant sea surface height indicative of the time-averaged streamlines. The model configuration of these simulations is identical to the reference simulation ($L_d/L_d^\text{ref} = 1$) under climatological wind and AMOC forcing, except for a decreased spacial resolution of roughly $1.4$° employed here to mimic the behaviour of low-resolution climate models. Compared to reference stratification (\textbf{a}) the GS is slightly faster close to the coast at strong stratification (\textbf{b} and \textbf{c}), as reported on in \citeA{peng2022surface}, but does not extend as an eastward jet in any stratification configuration.}
     \label{fig: low_res}
 \end{figure}

\begin{figure}[htbp!]
    \centering
    \includegraphics[width = \textwidth]{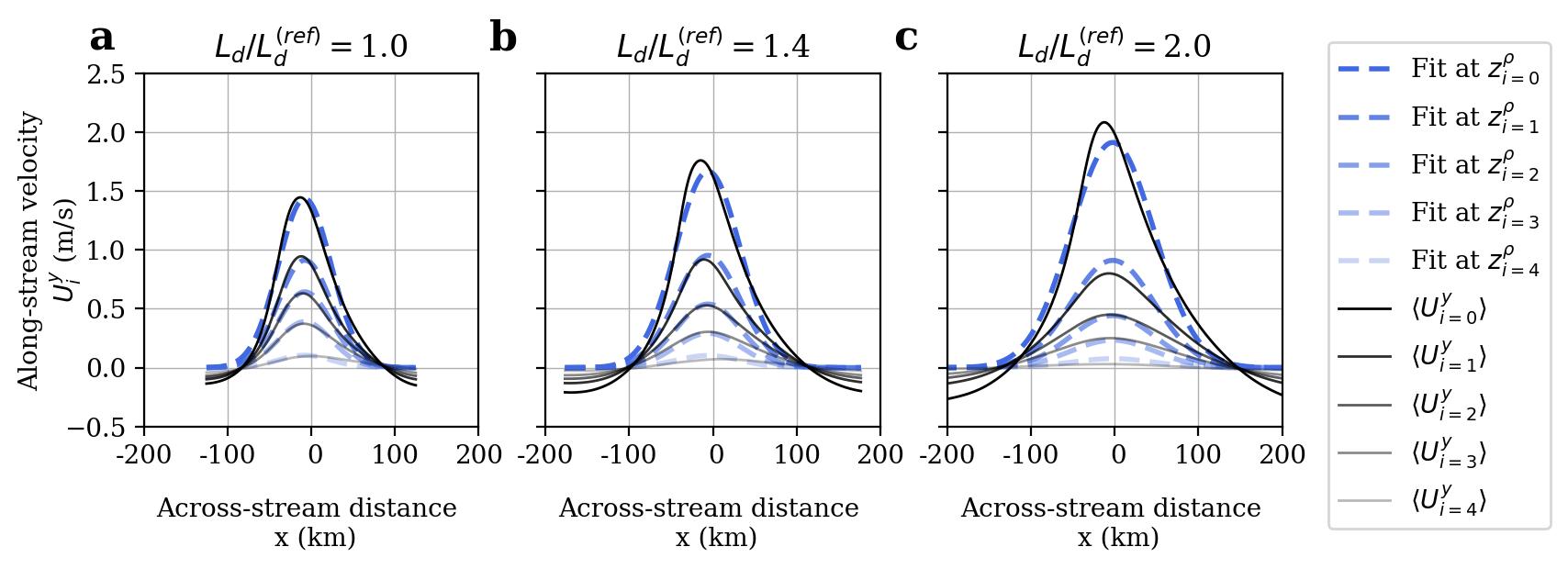}
    \caption{\textbf{Comparison between path averages of along-stream velocity $\mathbf{\langle U^y_i \rangle}$ and fitted profiles across all model layers (i = 0,...,4) for the GSE in the three stratification scenarios $\mathbf{L_d/L_d^\text{ref}}$ = 1 (a), 1.4 (b) and 2 (c) under climatological wind and AMOC forcing.} The details on the calculation of these profiles can be found in Supplementary Text S2. As stratification increases, the GSE becomes wider, accelerates at the surface and slows down at depth. In the core of the GSE, the velocity profiles $\langle U^y_i \rangle$ are well approximated by the fitted idealized profile given in equation~\ref{idealized_profile}. The fitted parameters are given in table~\ref{tab:fitted}.}
    \label{fig: velocity_profiles}
\end{figure}

\begin{figure}
     \centering
     \includegraphics[width = 0.45\textwidth]{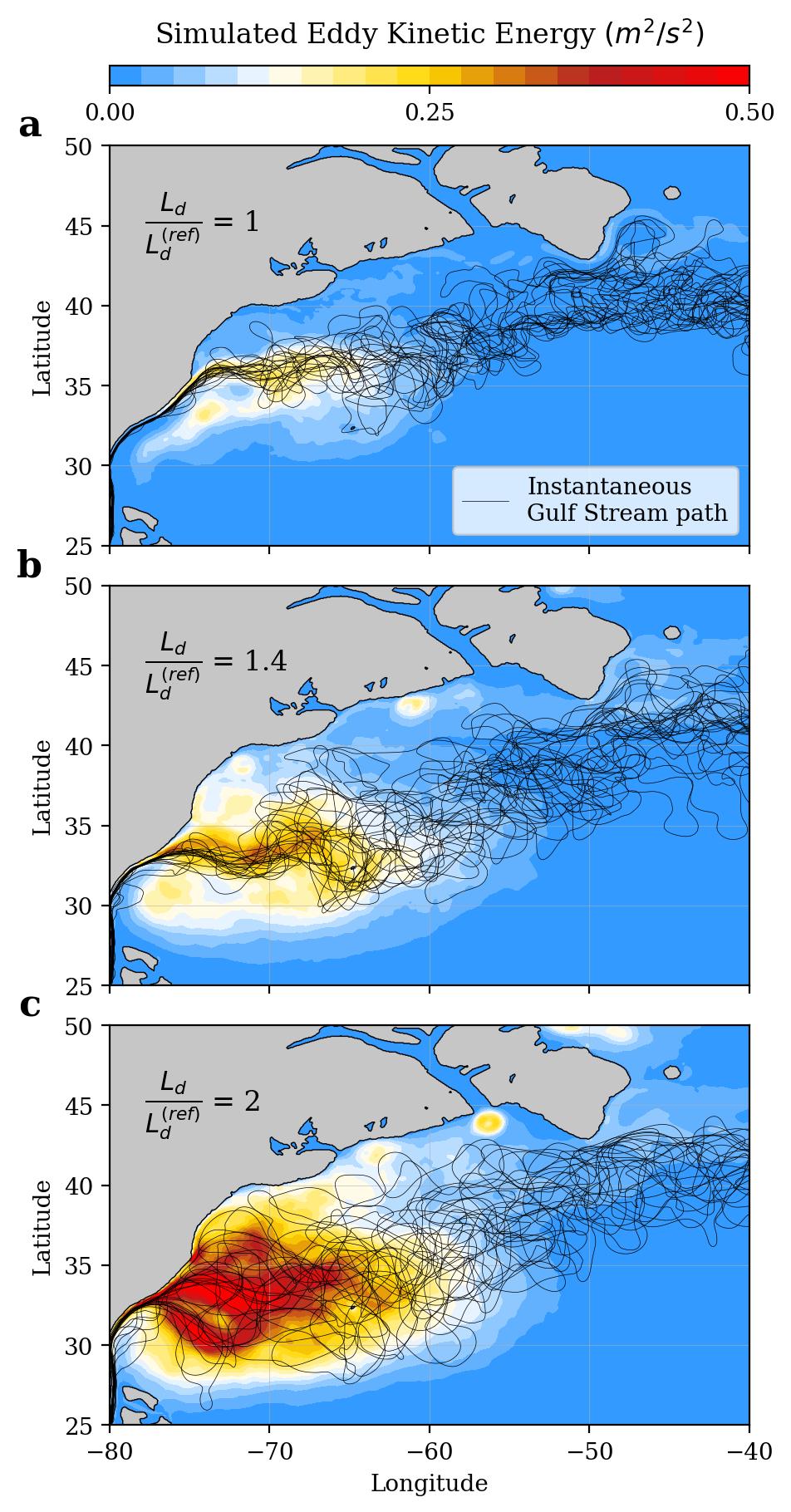}
     \caption{\textbf{EKE (color scale) and path lines of the GSE with climatological wind forcing but no AMOC at $\mathbf{L_d/L_d^{(\text{ref})}~=~1}$ (\textbf{a}), $\mathbf{L_d/L_d^{(\text{ref})}~=~1.4}$ (\textbf{b}) and $\mathbf{L_d/L_d^{(\text{ref})}~=~2}$ (\textbf{c})}. The displayed path lines are calculated in the same manner as for Fig.~2 of the main text. The shutdown of the AMOC is modeled by removing the mass fluxes at the zonal boundaries of the computational domain. The EKE signatures are weaker than with climatological AMOC forcing, but the amplifying response of path variability and EKE to increased stratification persists.}
     \label{fig: no_AMOC}
\end{figure}

\begin{figure}
     \centering
     \includegraphics[width = 0.45\textwidth]{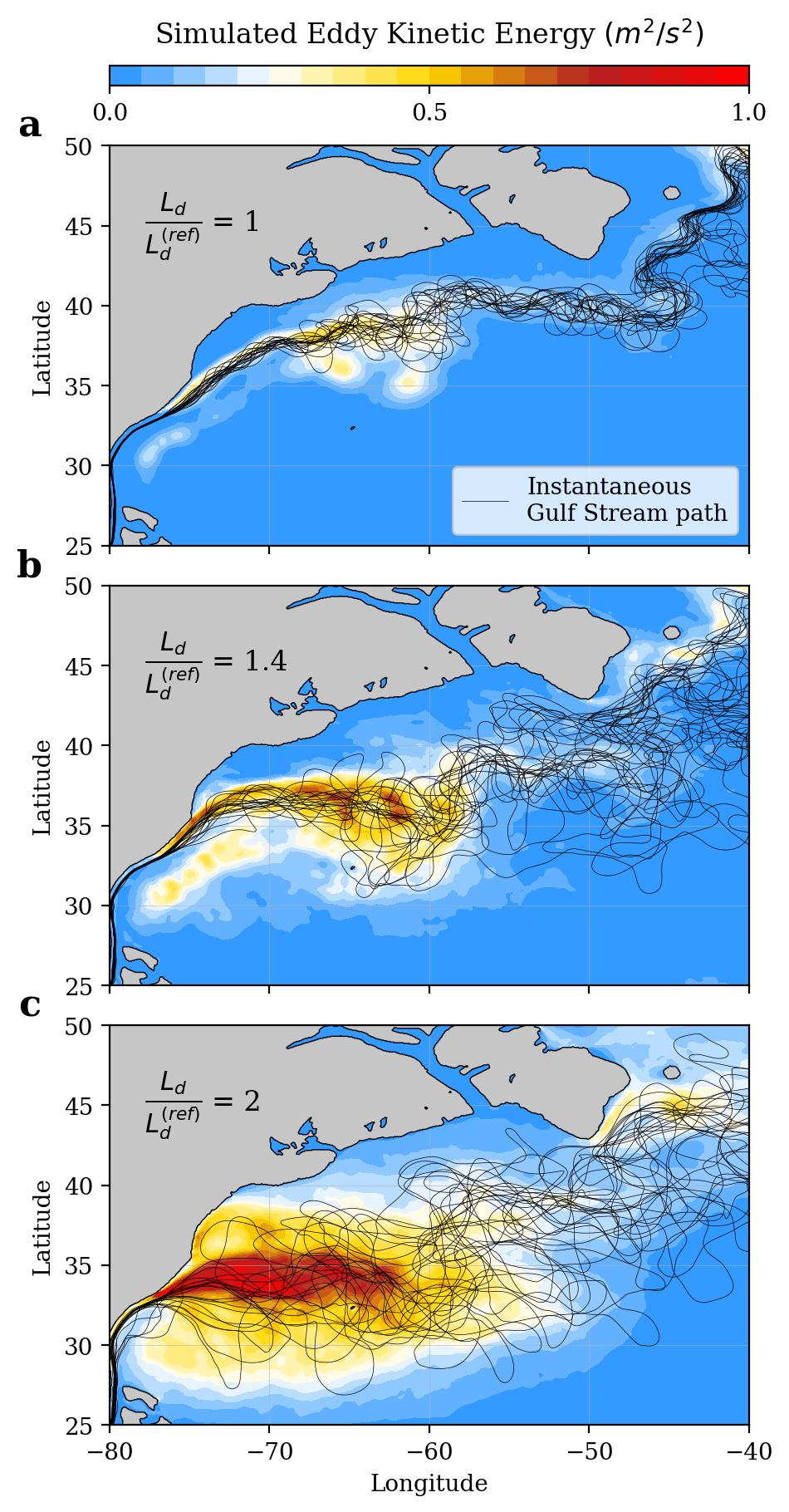}
     \caption{\textbf{EKE (color scale) and path lines of the GSE with a climatological AMOC and increased wind forcing at $\mathbf{L_d/L_d^{(\text{ref})}~=~1}$ (\textbf{a}), $\mathbf{L_d/L_d^{(\text{ref})}~=~1.4}$ (\textbf{b}) and $\mathbf{L_d/L_d^{(\text{ref})}~=~2}$ (\textbf{c})}. The displayed path lines are calculated in the same manner as for Fig.~2 of the main text. The increase of surface winds is modeled by multiplying the surface wind stress from \citeA{hellerman1983normal} by a factor of $1.5$. The EKE signatures are stronger than with climatological wind forcing, but the amplifying response of path variability and EKE to increased stratification persists.}
     \label{fig: strong_winds}
\end{figure}

\begin{figure}
     \centering
     \includegraphics[width = 0.45\textwidth]{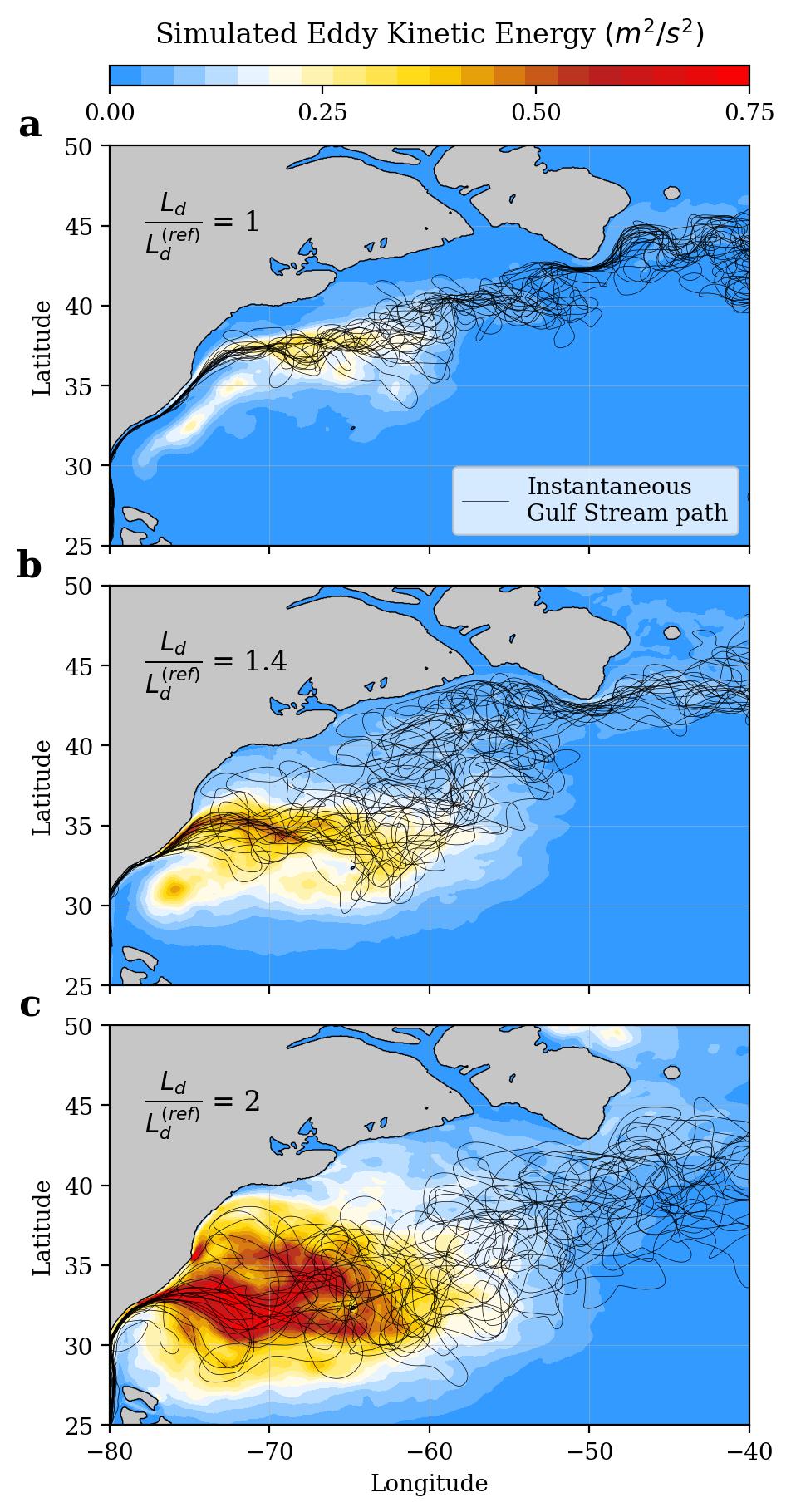}
     \caption{\textbf{EKE (color scale) and path lines of the GSE without AMOC forcing and increased wind forcing at $\mathbf{L_d/L_d^{(\text{ref})}~=~1}$ (\textbf{a}), $\mathbf{L_d/L_d^{(\text{ref})}~=~1.4}$ (\textbf{b}) and $\mathbf{L_d/L_d^{(\text{ref})}~=~2}$ (\textbf{c})}. The displayed path lines are calculated in the same manner as for Fig.~2 of the main text. The increase of surface winds is modeled by multiplying the surface wind stress from \citeA{hellerman1983normal} by a factor of $1.5$, and the AMOC shutdown is modeled by the removal of mass fluxes at the meridional boundaries of the domain. The amplifying response of path variability and EKE to increased stratification persists.}
     \label{fig: strong_winds_no_AMOC}
\end{figure}

\begin{figure}[htbp!]
    \centering
    \includegraphics[width = 0.8\textwidth]{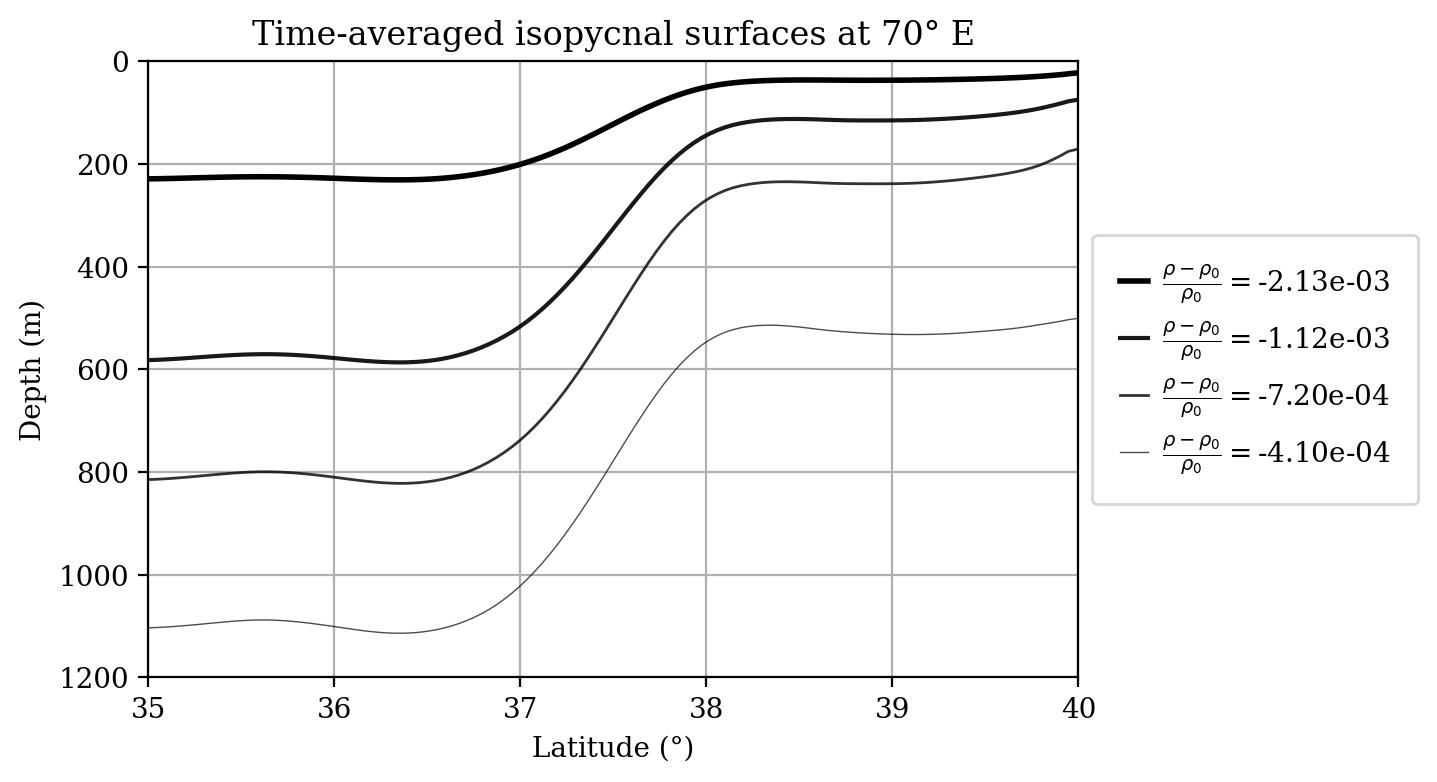}
    \caption{\textbf{Signature of Gulf Stream Extension in time-averaged heights of isopycnal surfaces in the reference model configuration ($\mathbf{L_d/L_d^\text{ref}=1}$) under climatological wind and AMOC forcing.} Through spatial variations of the isopycnal layer thicknesses, the model successfully reproduces the characteristic jump in thermocline depth across the Gulf Stream.}
    \label{fig: thermocline_jump}
\end{figure}

This supporting information includes model details (Text S1) and validation  against observations of surface flows (Fig.~\ref{fig: validation_AVISO+}) and AMOC transport (Fig.~\ref{fig: AMOC}), the methodology used to compute the density profiles of Fig.~1b and the deformation radius $L_d$ (Text S2), the employed methodology for calculating path averages of the moving GSE (Text S3 and Fig.~\ref{fig: velocity_profiles}), details on the stability analysis of the GSE (Text S4), low-resolution representations of the Gulf Stream (Fig.~\ref{fig: low_res}), amplifying variability response of the GSE to increased stratification under an AMOC shutdown (Fig.~\ref{fig: no_AMOC}), stronger surface winds (Fig.~\ref{fig: strong_winds}) and both an AMOC shutdown and stronger surface winds (Fig.~\ref{fig: strong_winds_no_AMOC}), the signature of the Gulf Stream Extension on the height of isopycnal layers in our model (Fig.~\ref{fig: thermocline_jump}) and a caption for a video of GSE variability under increased stratification (Movie S1).\\

\clearpage

\noindent\textbf{Text S1: Details and Validation of Circulation Model}\\

\vspace{10mm}
We employ a multi-layer isopycnal ocean circulation model whose dynamics are given by
\begin{align}
        \partial_t h_i + \mathbf{{\nabla}} \cdot \left( h_i \mathbf{u}_i \right)&=
        0, \label{mass_conservation}\\
        \partial_t \mathbf{u}_i + \mathbf{u}_i \cdot \nabla \mathbf{u}_i + \mathbf{f}\times\mathbf{u}_i&= - \nabla \phi_i + F + D \label{momentum_conservation}\\
        \frac{\partial \phi_i}{\partial z}&= -g\frac{\rho_i}{\rho_0} \label{hydrostasy}
\end{align}
Here $h_i$ denote the layer thicknesses, $\mathbf{u}_i = (u_i,v_i)$ denotes the horizontal fluid velocity, $\phi_i$ the dynamical pressure and $g$ the gravitational acceleration. The subscript $i$ denotes the layer index. The reference density is $\rho_0 = 1000$ kg/m$^3$ and $\rho_i$ are the layer densities. The Coriolis parameter is $\mathbf{f} = f \hat{z}$, and $F$ and $D$ contain further forcing and dissipative terms that will be described in the following. The system of equations \ref{mass_conservation}, \ref{momentum_conservation} and \ref{hydrostasy} also describes an Oberbeck-Boussinesq model that is formulated along isopycnal coordinates, and thus can be seen as an adiabatic version of the primitive equations. The basin-averaged layer thicknesses and densities are user-defined control parameters that set the basin-averaged stratification. Specifying these parameters therefore permits studying the response of the GSE to basin-averaged changes in stratification while still allowing for realistic deviations of stratification in space and time~\ref{fig: thermocline_jump}, as the local layer thicknesses $h_i$ are dynamic variables. 

We use a modern, GPU-enabled numerical implementation of an isopycnal ocean model \cite{popinet2020vertically} in a configuration that resembles the model of~\citeA{hurlburt2000impact}, who convincingly demonstrated that a limited number of isopycnal layers is sufficient to obtain realistic results for the surface and abyssal circulations in the North Atlantic provided a sufficient horizontal resolution is used. The underlying code can be accessed online (https://basilisk.fr/src/examples/gulf-stream.c, accessed on the 30th of June 2026) and has been verified in a hierarchy of test cases of increasing complexity. To further increase fidelity, we validated this model against observations and found that it realistically reproduces surface currents (Supplementary Fig.~S1) and the AMOC (Supplementary Fig.~S2).\\

The numerical domain is bounded by impermeable boundaries at constant longitude $\phi$ and latitude $\theta$ and extends across 9° N $<\theta<$ 51° N and 98° W $<\phi<$ 14° W. The simulations are performed on a numerical grid spanning 2048 grid points in longitude and 1024 grid points in latitude, resulting in a resolution of roughly 0.04° (close to 1°/25, or 5 km). We use five isopycnal layers, the upper four layers having an equal domain-averaged thickness of 250 m and the lowest layer fills the remaining volume to the ocean floor. The coastlines and topography are taken from the ETOPO2 dataset~\cite{noaa2022etopo}. The height of the layer interfaces are defined as 
\begin{equation}
    z_i = \eta + \sum_{j=0}^i h_j,
\end{equation}
with $\eta$ being the height of the topography. The sea surface height is simply $\text{SSH} = z_{i=5}$. The isopycnal heights are calculated as $z_i^\rho = (z_{i+1}+z_i)/2$. To prevent the thickness of isopycnal layers from vanishing \citeA{hurlburt2000impact} shrink the topography to remain confined to the lowermost layer, and the four upper layers are bounded by vertical side walls at their lateral boundaries. Although Basilisk can simulate the dynamics of fluid layers whose thickness vanishes, we shrink the topography as in~\citeA{hurlburt2000impact} in our numerical experiments. Enhanced bottom friction is applied at the location where topography is compressed in order to constrain the mass flux in the upper layers. The flow is forced by wind stress from the Hellermann climatology~\cite{hellerman1983normal}. For the simulations with stronger surface winds, the Hellermann wind stress is multiplied by a factor of 1.5. The cross-isopycnal flow is parameterized by entrainment formulas that allow for momentum transport across layers~\cite{hurlburt2000impact}. The AMOC is accounted for by adding mass fluxes at the meridional boundaries that represent a southward flowing deep western boundary current in the lower layers and a northward return flow in the surface layer. For the simulations without an AMOC, these mass fluxes are removed. All presented time-averaged quantities are taken over at least 5 years of simulation time.\\

Fig.~\ref{fig: validation_AVISO+} compares the surface circulation of the ocean model at reference stratification ($L_d/L_d^{ref}=1$, Supplementary Text S2) and under climatological wind and AMOC forcing to DUACS gridded satellite altimetry data~\cite{copernicus-marine-service_2025}. The model reproduces several important features of the subtropical gyre in the North Atlantic. Both instantaneous snapshots (Fig.~\ref{fig: validation_AVISO+}a, b) and time-averages (Fig.~\ref{fig: validation_AVISO+}c, d) of the sea surface height show that the signature of the gyre is on the order of $1$ m in both observations and the simulation. The GS detaches correctly in the simulation and retains it coherence until reattaching to the coast at the Grand Banks, thus correctly reproducing the mean path of the GS. Eddies are present in the simulation both close to the eastward jet and in the basin interior. Comparing Fig.~\ref{fig: validation_AVISO+}e and f shows that the model correctly reproduces the associated sea surface height variability present in observations.\\

The depression of sea surface height due to the subpolar gyre is less apparent in the simulation as in observations, and the contours of sea surface height close to the southern domain boundary are slanted more southwards in the observations than in the model. Both the northern depression of sea surface height and the slanting of isolines in the south are likely an artifact of the impermeable model domain boundaries, but should not affect the investigated model dynamics in the GSE.\\

The ocean circulation model also features a realistic representation of the AMOC, which we validate in Fig.~\ref{fig: AMOC} by comparing model meridional transport as a function of depth against observational data from the RAPID array \cite{moat2024atlantic}. The maximum model transport is slightly underestimated and decreases more strongly at depth compared to observations, but the model correctly reproduces the depth of maximum transport and the time variability of the AMOC.\\

\vspace{20mm}
\noindent\textbf{Text S2: Calculation of Stratification Profiles and $\mathbf{L_d}$}\\

\vspace{10mm}
The first baroclinic deformation radius $L_d$ is defined as the inverse of the smallest eigenvalue $\lambda$ of 
\begin{equation}
    \frac{\partial}{\partial z}\left( \frac{f^2}{N^2}\frac{\partial \psi}{\partial z}\right) = -\lambda^2 \psi,
\end{equation}
with boundary conditions on the eigenmodes $\psi$ being $\partial_z\psi=0$ at the surface ($z=0$) and the ocean floor ($z=\eta$). The stratification is characterized by the Brunt–Väisälä frequency, $N = \sqrt{-g/\rho_0 (\partial \rho/\partial z)}$. To compute $N$, we discretize $\partial\rho/\partial_z = (\rho_{i+1} - \rho_i)/(z_{i+1}^\rho - z_i^\rho)$, where $\rho_i$ and $z_i^\rho$ are the density and height of the $i^\text{th}$ isopycnal layer (Supplementary Text S1). A single representative value of $L_d$ is obtained by averaging values for $z_i^\rho$, $H$ and $f$ over the simulation domain. We solved this eigenvalue problem using a standard inverse solver~\cite{deremble_qgutils}.\\

The stratification for the reference runs ($L_d/L_d^{(\text{ref})} = 1$) follows~\citeA{hurlburt2000impact}. The model has a vertical resolution of five isopycnal layers with densities $\rho_i$. From bottom to top, the relative density anomalies in each of the layers are
\begin{equation}
   \frac{\rho_i - \rho_0}{\rho_0} = [0, -0.41, -0.72, -1.12, -2.13] \times 10^{-3}.\label{isopycnal_densities}
\end{equation}
To simulate the effects of increased ocean stratification these reference values of $(\rho_i - \rho_0)/\rho_0$ are multiplied by $2$ and $4$. This yields characteristic deformation radii of $L_d = 32.0$ km ($L_d/L_d^{(\text{ref})} = 1$),  $45.2$ km ($L_d/L_d^{(\text{ref})} = 1.4$) and $64.1$ km ($L_d/L_d^{(\text{ref})} = 2$) for the numerical experiments.\\

To underline the relevance of these simulations we provide an estimate of stratification in the GS region today and by the end of the 21$^{\text{st}}$ century in Fig.~1b and compare it to the density profiles in our simulations. To this end we first calculate a current GS stratification climatology $\rho^{ref}_{GS}(z)$ using the Levitus dataset~\cite{levitus1982climatological}. In order for the calculated profile to be a reference stratification with respect to the observed trends in \citeA{todd2023warming}, the GS region is defined as the polygon spanned by the four points 
\begin{equation}
    [(80\text{°W}, 26\text{°N}), (80\text{°W}, 32\text{°N}), (\text{68}\text{°W}, \text{38}\text{°N}), (\text{68}\text{°W}, \text{44}\text{°N})].\label{polygon}
\end{equation}
The resultant profile, displayed in Fig.~1b, is well approximated by the model stratification under reference conditions ($L_d/L_d^{(\text{ref})} = 1$). We then estimate the future density profile in the GS region $\rho^{proj}_{GS}(z)$ based on a linear extrapolation of the observed surface trends $d\rho/dt(z = 0)$ as
\begin{equation}
    \rho^{proj}_{GS}(z) = \rho^{ref}_{GS}(z) \ + \ \frac{d\rho}{dt}(z=0)\times\frac{\rho_{GS}^{ref}(z) - \rho_{GS}^{ref}(z = \eta)}{\rho_{GS}^{ref}(z=0) - \rho_{GS}^{ref}(z = \eta)}\times t^{proj},
\end{equation}
where $t^{proj}$ is the projection time. The estimation uses only the observed surface trends, based on the the similar vertical structure of both the density profile ($\rho^{ref}_{GS}(z)$ in Fig.~1b) and the observed trends (Fig.~3 of \citeA{todd2023warming}) in the Gulf Stream region. Fig.~1b shows an estimation of the GS stratification by the end of the 21$^{\text{nd}}$ century based on the Levitus reference profile (\citeA{levitus1982climatological}, $t^{proj} = 2100 - 1982 = 118$ yr) and the currently observed surface trend $d\rho/dt(z=0)$ =~$-0.015$~$(\pm0.005)$~kg~m$^{-3}$~yr$^{-1}$ (see Fig.~3 of \citeA{todd2023warming}). The density profiles for our simulations are calculated by averaging the isopycnal heights $z_i^\rho$ across the area defined by the vertices in~\ref{polygon}. The resulting isopycnal heights are then plotted against the density anomalies in Fig.~1b. The simulation at $L_d/L_d^{(\text{ref})} = 1.4$ approximates the projected density profile reasonably well.\\

\vspace{20mm}
\noindent\textbf{Text S3: Stream-wise Averaging of the GSE}\\

\vspace{10mm}
We here define stream-wise averages based on the GSE path, with its geographical coordinates denoted as $\phi_\text{GSE}(y), \theta_\text{GSE}(y)$. In our model the GSE path coincides with the 0m contour line of the sea surface height (SSH$(\phi_\text{GSE}, \theta_\text{GSE}) = 0m$). The along-stream position $y$ denotes the arclength of the GSE path obtained by integrating
\begin{equation}
    dy = \sqrt{d\Tilde{x}^2 + d\Tilde{y}^2} \ \text{with} \  (d\Tilde{x}, d\Tilde{y}) = \mathcal{S} \cdot (d\phi, d\theta).
\end{equation}
Here, $(d\Tilde{x}, d\Tilde{y})$ are the cartesian differentials along the direction of the geographical basis vectors $\boldsymbol{\hat{\phi}}, \boldsymbol{\hat{\theta}}$ and $\mathcal{S}$ is the operator that transforms geographical differentials to cartesian differentials in the GSE core, given by
\begin{equation}
 \mathcal{S} = \begin{bmatrix}
  \frac{2\pi R \cos{\theta_\text{GSE}(y)}}{360\text{°}} & 0 \\
  0 & \frac{2\pi R}{360\text{°}}
\end{bmatrix}.
\end{equation}
The integration of $dy$ is carried out downstream of the separation point, which itself is defined as the point on the GSE path that is closest to $\phi$ = 75 °E, $\theta$ = 35 °N.\\

To calculate the GSE path variability $s_\text{GSE}$ of equation (1) in the main text we calculate path averages as 
\begin{equation}
    \langle \theta_\text{GSE} \rangle^\phi = \frac{\iint \ \theta_\text{GSE} \  dy \ dt}{\iint \ dy\ dt},
\end{equation}
where the integration in y is carried out first across segments of the GSE path where $\phi < \phi_\text{GSE} < \phi + 1$°, and the integration in time is carried out second across all available simulation snapshots. \\
We now wish to define a distance $x$ that holds the across-stream position relative to the core of the GSE. We define the unit vectors of the stream-wise coordinates as
\begin{equation}
    \hat{\mathbf{y}} = \left(\cos(\alpha), -\sin(\alpha)\right) \ , \ \hat{\mathbf{x}} = \left(\sin(\alpha), -\cos(\alpha)\right),
\end{equation}
where $\hat{\mathbf{y}}$ defines the along-stream direction, $\hat{\mathbf{x}}$ defines the across-stream direction and $\alpha$ denotes the anticlockwise angle between the northward unit vector and the GSE path. It is calculated as
\begin{equation}
    \alpha = \arctan\left(-\frac{\Tilde{x}'}{\Tilde{y}'}\right)
\end{equation}
where $(\Tilde{x}', \Tilde{y}') = \mathcal{S}^{-1} \cdot (\phi_\text{GSE}', \theta_\text{GSE}')$. The mapping $\mathcal{M}$ between geographical ($\phi$, $\theta$) and stream-wise coordinates ($x$,$y$) is then defined by
\begin{equation}
    (\phi, \theta) = \mathcal{M}(x,y) = (\phi_\text{GSE}(y), \theta_\text{GSE}(y)) + x \left(\mathcal{S}(y)^{-1}\cdot\hat{\mathbf{x}}(y)\right). \label{transform}
\end{equation}
We carried out stream-wise averages along the GSE path for the simulations with climatological wind and AMOC forcing. The stream-wise average of the along-stream velocity $U^y_i = \mathbf{u}_i.\mathbf{\hat{y}}$, where $i$ is the layer index, is calculated as
\begin{equation}
    \langle U^y_i \rangle (x) = \frac{\iint \ U^y_i \  dy \ dt}{\iint \ dy \ dt}. 
\end{equation}
The differentials $dy(x,y)$ are calculated numerically at each snapshot with the transform in~\ref{transform}. The resulting stream-wise averaged profiles of $U^y_i$ are displayed in figure \ref{fig: velocity_profiles}. The averages of topography ($\eta$), the topographical gradients ($\partial_x\eta, \partial_y\eta$), the Coriolis parameter ($f$), its gradients ($\partial_xf, \partial_yf$) and the layer thicknesses ($H_i$) are calculated with an additional average in the x-direction, 
\begin{equation}
    \langle f \rangle = \frac{\iiint \ f \ dy \ dt \ dx}{\iiint \ dy \ dt \ dx}.
\end{equation}
Satisfyingly converged averages are obtained when along-stream integration is carried out between $0\text{m} < y < 2000 \text{km}$ and the across-steam integrals between $-x_b < x < x_b$, with increasingly large across-stream limits ($x_b$ = $125$ km, $177$ km, $250$ km for $L_d/L_d^{ref}$ = 1, 1.41, 2) to account for the widening of the GSE. The results are summarized in table \ref{tab:streamaverages}.\\

\vspace{20mm}
\noindent\textbf{Text S4: Linear Stability Analysis of the Gulf Stream Extension}\\

\vspace{10mm}
A linear stability analysis is presented as a dynamical rationale underlying the amplified variability of the GSE path for the three stratification scenarios ($L_d/L_d^\text{ref} = 1, 1.4, 2$) under climatological wind and AMOC forcing. We aim to understand changes in the GSE dynamics away from the coastlines, and therefore investigate the stability of the GSE under a locality assumption~\cite{dimas1995baroclinic} as a current whose velocity profile is initially invariant in the stream-wise direction. The temporal evolution of perturbations of the GSE is then modeled with a five-layer quasi-geostrophic model~\cite{vallis2017atmospheric}.\\

We here adopt a notation where layer indices $i$ increase from bottom to top. The coordinates of our quasi-geostrophic model denote the across-stream ($x$) and along-stream ($y$) position, with vector components written as $\mathbf{u}_i = (u^x_i, u^y_i)$. The model dynamics are given by
\begin{equation}
    \partial_tq_i + J(\psi_i, q_i) = \nu \nabla^2\omega_i.\label{QG}
\end{equation}
The streamfunction $\psi_i$ is related to the across-stream velocity as $u^x_i = -\partial_y\psi_i$ and to the along-stream velocity as $u^y_i = \partial_x\psi_i$. The Jacobian is defined as $J(A,B) = \partial_xA \ \partial_y B - \partial_yA \ \partial_x B$. We include a viscous dissipation of the relative vorticity $\omega_i = \nabla^2\psi_i$ with $\nu = 100$m$^2$s$^{-1}$ to suppress slowly growing submesoscale modes~\cite{lacasce2019jet}, and have confirmed that this does not affect the presented maximum growth rates at the scales of the deformation radius. The potential vorticity $q_i$ in the interior layers ($i = $1,2,3) is given by
\begin{equation}
    q_i = \nabla^2\psi_i + \frac{f_0^2}{H_i}\left(\frac{\psi_{i+1} - \psi_i}{g'_{i+1}} - \frac{\psi_{i} - \psi_{i-1}}{g'_{i}}\right) + f.\label{PVdef}
\end{equation}
Here $H_i$ are the layer thicknesses, $f_0$ the reference Coriolis parameter and $g'_i = g(\rho_{i-1} - \rho_{i})/\rho_0$ is the reduced gravity. In the outer layers ($i=$0,4), $q_i$ is given by
\begin{eqnarray}
     q_0 &=& \nabla^2\psi_0 + \frac{f_0^2}{H_0}\left(\frac{\psi_{1} - \psi_0}{g'_{1}}\right) + f + \frac{f_0}{H_0}\eta, \label{PVdef_lower}\\
     q_4 &=& \nabla^2\psi_4 + \frac{f_0^2}{H_4}\left(\frac{\psi_{3} - \psi_{4}}{g'_{4}}\right) + f.\label{PVdef_upper}
\end{eqnarray}
Here, $\eta$ is the height of the topography. 
In our quasi-geostrophic model, both the Coriolis parameter $f$ and the topography $\eta$ can vary linearly as functions of $x$ and $y$ ($f = f_0 + \partial_xf \ x + \partial_yf \ y$, $\eta = \eta_0 + \partial_x\eta \ x + \partial_y\eta \ y$).\\

The quasi-geostrophic model is calibrated to resemble the dynamical conditions of the GSE in the comprehensive model runs. The parameters $\eta_0$, $\partial_x\eta$, $\partial_y\eta$, $f_0$, $\partial_xf$, $\partial_yf$ and $H_i$ are calculated as path averages of equivalent model parameters along the moving GSE path (Supplementary Text S3, Table S1). The values of the reduced gravity $g_i'$ were calculated directly from the layer densities employed in each stratification scenario.\\

In the stability analysis the GSE is approximated by the idealized velocity profile
\begin{equation}
    U^y_i(x) = \left(\Delta U e^{\frac{z^\rho_i}{\delta_v}} + U_{0}\right)\ e^{-\frac{(x - x_0)^2}{\delta_h^2}}.\label{idealized_profile}
\end{equation}
Here, $x_0$ is the offset between the GSE path and the velocity maximum, $U_{0}$ is the velocity in the abyss, $\Delta U$ is the vertical velocity deficit and $\delta_h$, $\delta_v$ are the horizontal and vertical decay scales of the profile, respectively. These parameters are obtained by fitting equation~\ref{idealized_profile} against path averages of the along-stream velocity (Supplementary Text S3). The results can be consulted in table~S2 of the supplementary material. Supplementary Fig.~S4 shows that the fitted idealized profiles provide good approximations of the path averaged profile of the GSE.\\

The stability analysis of the GSE is carried out using standard procedures. We decompose the vorticity $q= q'_i + Q_i$ into small perturbations $q_i'$ around a base state vorticity $Q_i$. To calculate the base state $Q_i$, we integrate $\partial_x\Psi = U_i^y$ with $U_i^y$ given by~\ref{idealized_profile} to obtain the base state streamfunction $\Psi_i$ which is then used to compute~\ref{PVdef}, \ref{PVdef_lower} and~\ref{PVdef_upper}. An evolution equation of $q_i'$ is obtained by linearizing~\ref{QG} around $Q_i$,
\begin{equation}
    \partial_tq'_i + J(\psi_i', Q_i) + J(\Psi_i, q'_i) = \nu \nabla^2\omega'_i.\label{perturbations}
\end{equation}
The perturbation vorticity $q'_i$ is calculated from the perturbation stream function $\psi'_i$ without the Coriolis and topography terms in~\ref{PVdef}, \ref{PVdef_lower} and~\ref{PVdef_upper}.\\

The stream-wise invariance of $U^y_i$ permits a decomposition of the dynamics into Fourier modes in time and along the $y$-direction. We therefore investigate the evolution of perturbation modes of the form
\begin{equation}
    \psi_i' = \operatorname{Re}\left(\Phi(x)\ e^{i\left(l y - \sigma t\right)}\right),\label{mode}
\end{equation}
where $l$ is the along-stream wavenumber, $\sigma$ the complex frequency and $\operatorname{Re}$ denotes the real part of the bracketed expression. The expression~\ref{mode} is inserted into~\ref{perturbations} and the resulting eigenvalue problem for $\sigma$ is solved numerically after discretizing the spacial derivatives in the cross-stream direction. The python scripts that carry out this analysis are linked in the open research section of this article.\\

\vspace{20mm}
\noindent\textbf{Movie S1. Visualization of increased Gulf Stream Extension Path Variability.} \\

\vspace{10mm}
4-year time series of surface flow speed $|\mathbf{u}|$ for reference ($L_d/L_d^{(\text{ref})}~=~1$) and increased stratification ($L_d/L_d^{(\text{ref})}~=~1.4\,,2$) scenarios under climatological wind and AMOC forcing. Increased stratification leads to larger meanders of the GSE and an increased path variability (quantified in Fig.~3a). The movie also illustrates that the GSE becomes faster under increased stratification and sheds increasingly larger and more energetic rings, which is reflected in the increase in EKE displayed in Fig.~3b.\\


\clearpage

\renewcommand{\arraystretch}{1.5}
\begin{table}
\begin{center}
\begin{tabular}{|C{4cm}||C{3cm}|C{3cm}|C{3cm}|}
\hline
Parameter & $L_d/L_d^{ref} = 1$ & $L_d/L_d^{ref} = 1.4$ & $L_d/L_d^{ref} = 2$ \\ 
\hline \hline
$\langle f \rangle$ (s$^{-1}$)
& $8.9\times10^{-5}$ & $8.7\times10^{-5}$ & $8.2\times10^{-5}$ \\
\hline
$\langle \partial_x f \rangle$ (m$^{-1}$s$^{-1}$)
& $-1.3\times10^{-11}$ & $-1.2\times10^{-11}$ & $-1.2\times10^{-11}$ \\
\hline
$\langle \partial_y f \rangle$ (m$^{-1}$s$^{-1}$)
& $5.0\times10^{-12}$ & $2.2\times10^{-12}$ & $-2.0\times10^{-14}$ \\
\hline
$\langle \eta \rangle$ (m)
& $-4.8\times10^{3}$ & $-4.7\times10^{3}$ & $-5.0\times10^{3}$ \\
\hline
$\langle \partial_x \eta \rangle$
& $-3.5\times10^{-3}$ & $-3.6\times10^{-3}$ & $-1.6\times10^{-3}$ \\
\hline
$\langle \partial_y \eta \rangle$
& $1.2\times10^{-3}$ & $1.4\times10^{-3}$ & $1.1\times10^{-3}$ \\
\hline
$\langle H_0\rangle$ (m)
& 3760 & 3659 & 3979 \\
\hline
$\langle H_1 \rangle$ (m)
& 380 & 294 & 244 \\
\hline
$\langle H_2\rangle$ (m)
& 191 & 283 & 265 \\
\hline
$\langle H_3\rangle$ (m)
& 174 & 175 & 233 \\
\hline
$\langle H_4\rangle$ (m)
& 276 & 245 & 242 \\
\hline
\end{tabular}
\end{center}

\vspace{10mm}

\caption{Stream-wise averages of the Coriolis parameter $f$, its gradients $\partial_xf$, $\partial_yf$, the topography $\eta$, its gradients $\partial_x\eta$, $\partial_y\eta$ and the layer thicknesses $H_i$ for the three simulations with climatological AMOC and wind forcing under increasingly strong stratification scenarios ($L_d/L_d^{ref} = 1$, $1.4$ and $2$).}
\label{tab:streamaverages}
\end{table}

\renewcommand{\arraystretch}{1.5}
\begin{table}
\begin{center}
\begin{tabular}{|C{3.5cm}||C{3cm}|C{3cm}|C{3cm}|}
\hline
Parameter & $L_d/L_d^{ref} = 1$ & $L_d/L_d^{ref} = 1.4$ & $L_d/L_d^{ref} = 2$ \\
\hline \hline
$\Delta U$ (m/s)
& $1.81$ & $2.24$ & $2.74$ \\
\hline
$U_0$ (m/s)
& $0.10$ & $0.10$ & $0.08$ \\
\hline
$x_0$ (m)
& $-7.5\times10^{3}$ & $-5.8\times10^{3}$ & $-2.4\times10^{3}$ \\
\hline
$\delta_h$ (m)
& $4.5\times10^{4}$ & $5.5\times10^{4}$ & $7.0\times10^{4}$ \\
\hline
$\delta_v$ (m)
& $454$ & $346$ & $303$ \\
\hline
\end{tabular}
\end{center}

\vspace{10mm}
\caption{Values of fitted parameters for the approximated velocity profile used in the linear stability analysis of the GSE (Supplementary Text S4) for the three stratification scenarios under climatological wind and AMOC forcing. The values are obtained by fitting the profile in~\ref{idealized_profile} to path averages of the along-stream GSE velocity $\langle U^y_i \rangle$ (Supplementary Text S3). Fig~\ref{fig: velocity_profiles} shows that the fits provide a good approximation for the path-averaged velocity profiles of the GSE.}
\label{tab:fitted}
\end{table}

\clearpage


\end{document}